\input phyzzx.tex
\tolerance=1000
\voffset=-0.0cm
\hoffset=0.7cm
\sequentialequations
\def\rl{\rightline}

\def\t1{{\tilde 1}}

\def\t{\theta}

\REF{\CON}{G. Aldazabal, L. E. Ibanez, F. Quevedo and A. M. Uranga, JHEP {\bf 0008} (2000) 002, hep-th/0005067; J. F. G. Cascales, M. P. Garcia del Moral, F. Quevedo and A. M. Uranga, 
hep-th/0312051.}
\REF{\DSS}{G. Dvali, Q. Shafi and S. Solganik, hep-th/0105203.}
\REF{\BUR}{C. P. Burgess at. al. JHEP {\bf 07} (2001) 047, hep-th/0105204.}
\REF{\ALE}{S. H. Alexander, Phys. Rev {\bf D65} (2002) 023507, hep-th/0105032.}
\REF{\EDI}{E. Halyo, hep-ph/0105341.}
\REF{\SHI}{G. Shiu and S.-H. H. Tye, Phys. Lett. {\bf B516} (2001) 421, hep-th/0106274.}
\REF{\KAL}{C. Herdeiro, S. Hirano and R. Kallosh, JHEP {\bf 0112} (2001) 027, hep-th/0110271.}
\REF{\BEL}{J. Garcia-Bellido, R. Rabadan and F. Zamora, JHEP {\bf 01} (2002) 036, hep-th/0112147.}
\REF{\CAR}{K. Dasgupta, C. Herdeiro, S. Hirano and R. Kallosh, Phys. Rev. {\bf D65} (2002) 126002, hep-th/0203019.}
\REF{\JST}{N. T. Jones, H. Stoica and S. H. Tye, JHEP {\bf 0207} (2002) 051, hep-th/0203163.}
\REF{\LAS}{E. Halyo, hep-th/0307223.}
\REF{\BLO}{E. Halyo, hep-th/0312042.} 
\REF{\HYB}{A. D. Linde, Phys. Lett. {\bf B259} (1991) 38; Phys. Rev. {\bf D49} (1994) 748, astro-ph/9307002.} 
\REF{\MAP}{C. L. Bennett et. al., astro-ph/0302207; G. Hinshaw et. al., astro-ph/0302217; A. Kogut et. al., astro-ph/0302213.}
\REF{\HKO}{C. P. Herzog, I. R. Klebanov and P. Ouyag, hep-th/0205100.}
\REF{\KLE}{I. R. Klebanov and E. Witten, Nucl. Phys. {\bf B536} (1998) 199, hep-th/9807080.}
\REF{\MAL}{J. Maldacena and C. Nunez, Phys. Rev. Lett. {\bf 88} (2001) 588, hep-th/0008001; Int. Journ. Mod. Phys. {\bf A16} (2001) 822, hep-th/0007018.}
\REF{\BER}{M. Bertolini, hep-th/0303160 and references therein.} 
\REF{\ZAF}{F. Bigazzi, A. L. Cotrone, M. Petrini and A. Zaffaroni, hep-th/0303319 and references therein.}
\REF{\ACH}{J. Urrestilla, A. Achucarro and A. C. Davis, hep-th/0402032; A. Achucarro and T. Vachaspati, hep-th/9904229; A.Achucarro, A. C. Davis, M. Pickles and J. Urrestilla, hep-th/0109097.}
\REF{\DGM}{M. R. Douglas, B. R. Greene and D. R. Morrison, Nucl. Phys. {\bf B506} (1997) 84, hep-th/9704151.}
\REF{\DTE}{E. Halyo, Phys. Lett. {\bf B387} (1996) 43, hep-ph/9606423.} 
\REF{\BIN}{P. Binetruy and G. Dvali, Phys. Lett. {\bf B450} (1996) 241, hep-ph/9606342.}
\REF{\TYP}{E. Halyo, Phys. Lett. {\bf B454} (1999) 223, hep-ph/9901302.}
\REF{\TYE}{S. Sarangi and S. H. Tye, Phys. Lett. {\bf B536} (2002) 185, hep-th/204074; N. T. Jones, H. Stoica and S. H. Tye, Phys. Lett. {\bf B563} (2003) 6, hep-th/0303269;
G. Dvali and A. Vilenkin, hep-th/0312007.}
\REF{\POL}{E. J. Copeland, R. C. Myers and J. Polchinski, hep-th/0312067.}
\REF{\COS}{G. Dvali, R. Kallosh and A. Van Proeyen, hep-th/0312005.}
\REF{\STR}{E. Halyo, hep-th/0312268.}
\REF{\QUI}{E Halyo, JHEP {\bf 0110} (2001) 025, hep-ph/0105216.}
\REF{\STR}{A. Strominger, JHEP {\bf 010} (2001) 034, hep-th/0106113.}
\REF{\INF}{A. Strominger, hep-th/0110087.}
\REF{\HOL}{E. Halyo, hep-th/0203235.}

\singlespace
\rl{hep-th/0402155}
\rl{\today}
\pagenumber=0
\normalspace
\medskip
\bigskip
\titlestyle{\bf{D--Brane Inflation on Conifolds}}
\smallskip
\author{ Edi Halyo{\footnote*{e--mail address: vhalyo@stanford.edu}}}
\smallskip
 \centerline{California Institute for Physics and Astrophysics}
\centerline{366 Cambridge St.}
\centerline{Palo Alto, CA 94306}
\smallskip
\vskip 2 cm
\titlestyle{\bf ABSTRACT}

We describe a model of D--brane inflation on fractional D3 branes transverse to a resolved and deformed conifold. The resolution and the deformation are both necessary for inflation.
The fractional branes slowly approach each other along the $S^3$ and separate along the $S^2$ in the base of the conifold. We show that on the brane this corresponds to hybrid inflation. We 
describe the model also in terms of intersecting branes.

\singlespace
\vskip 0.5cm
\endpage
\normalspace

\centerline{\bf 1. Introduction}
\medskip

The idea that we live on a three--dimensional brane is very intriguing and has become very popular in recent years. There are a number of semi-realistic realizations of this idea 
in string theory which use D--branes. A particularly simple construction is in terms of D3 branes which are transverse to a conifold singularity of a six--dimensional Calabi--Yau 
manifold[\CON]. 
In this letter, we show how cosmological inflation can occur in such models. The type of inflation is necessarily D--brane inflation realized by
two branes slowly moving towards each other in the bulk[\DSS-\LAS]. In particular, we consider a variant of this scenario in which inflation occurs on fractional branes and arises 
due to the resolution of a singularity in the geometry transverse to the brane. (For the first example of this see ref. [\BLO].)

We show that inflation occurs when we consider fractional D3 branes on a resolved and deformed conifold. A conifold can be described as a cone over $S^2 \times S^3$.
We consider a D3 brane transverse to the conifold separated into two fractional D3 branes on the $S^3$. The origin
of inflation is the resolution of the conifold singularity which replaces the tip of the cone with an $S^2$ of finite size. The inflaton mass arises from another deformation of the conifold
towards an $ALE \times T^2$ (which is basically a deformation of the fiber structure) and results in a slow motion of the two fractional branes towards each other. On the brane world--volume, 
this corresponds to hybrid inflation[\HYB]. We obtain a brane/bulk
dictionary for all the parameters and show that the WMAP constraints[\MAP] can be satisfied for the string parameters. We also describe the relation of our model to other D--brane inflation 
models which are given in terms of Hanany--Witten constructions[\KAL,\LAS]. 

This letter is organized as follows. In the next section, we describe the conifold and its resolution and deformation that are relevant for our purposes. In section 3, we obtain the
scalar potential on the D3 brane world--volume and show that it leads to hybrid inflation that satisfies the WMAP constraints. In section 4 we give an alternative description of our model 
in terms of Hanany--Witten constructions. Section 5 includes a discussion of our results and our conclusions.

\bigskip
\centerline{\bf 2. The Conifold and Its Deformations}
\medskip

In this section we describe the conifold and two of its deformations. (Throughout the paper we use the term deformation for deviations of the conifold from its standard form and not
for the blow--up of an $S^3$ at its tip.[\HKO]) We describe two different deformations of the conifold: its resolution by an $S^2$ at the tip and a deformation
of the conifold towards $ALE \times T^2$. As we will see in the next section, these two deformations will give us the anomalous D--term and the inflaton mass on the world--volume
of a D3 brane transverse to the conifold.

The conifold is defined by the equation
$$z_1^2+z_2^2+z_3^2+z_4^2=0 \eqno(1)$$
which can also be written as
$$z_1^2+z_2^2+z_3^2=-z_4^2 \eqno(2)$$
In this form, the conifold is described as a fibration where the base is the $z_4$ plane (or the compactified sphere)
and the fiber is a $Z_2$ ALE space with a size that varies linearly with $|z_4|$. Let us define $z_4=X_4+iX_5$ and take the ALE space to be along the $X_6,X_7,X_8,X_9$ directions. 
In the orbifold limit, the ALE space is defined by the $Z_2$ identification $(X_6,X_7,X_8,X_9)=-(X_6,X_7,X_8,X_9)$. Thus there is a fixed plane (given by $z_4$) at 
$X_6=X_7=X_8=X_9=0$. In addition, the size of the whole ALE space vanishes at $X_4=X_5=0$ which gives the conifold singularity. 

The conifold can also be described as a cone over the space $T^{1,1}$ with the metric
$$ds^2=dr^2+r^2ds^2_{T^{1,1}} \eqno(3)$$
with
$$ds^2_{T^{1,1}}={1 \over 9}(d\psi+ \sum_{i=1}^2 cos \theta_i d \phi_i)^2+ {1\over 6} \sum_{i+1}^2 (d \theta_i^2+ sin^2 \theta_i d \phi_i)^2 \eqno(4)$$
as the metric for $T^{1,1}$. Here $(\theta_1, \phi_1)$ and $(\theta_2, \phi_2)$ parametrize two $S^2$s. We see that the space can be seen as a $S^1$ (or $U(1)$) bundle fibered over
$S^2 \times S^2$. Since an $S^1$ fibered over an $S^2$ gives an $S^3$, the conifold is a cone over $S^2 \times S^3$. At the tip of the cone both spheres shrink to zero size.
In terms of the coordinates $X_i$, the (compactified) planes $X_{4,5}$ and $X_{8,9}$
describe the two $S^2$s. The $S^1$ fiber is along the $X_6$ direction whereas the transverse directon parametrized by $r$ is along $X_7$. 

The first deformation of the conifold we are interested in is its resolution. This is obtained by replacing the singular tip of the cone or the orbifold singularity of the ALE fiber
by an $S^2$. The $Z_2$ ALE fiber at $z_4=0$ is given by
$$z_1^2+z_2^2+z_3^2=0 \eqno(5)$$
We can resolve the singularity by blowing up the a sphere of radius $R$. This is described by
$$z_1^2+z_2^2+z_3^2=R^2 \eqno(6)$$
If we write $z_i=x_i+iy_i$ then eq. (6) becomes
$$x_i^2-y_i^2=R^2 \qquad \qquad x_i y_i=0 \eqno(7)$$ 
Now define $r^2=x_i^2+y_i^2$. Then for $r^2=R^2$ we need to have $y_i=0$ and $x_i^2=R^2$ which is a sphere of radius $R$. After the resolution, the ALE 
space becomes a smooth Eguchi--Hanson space. The above sphere replaces the singular tip of the cone in eq. (2); at the tip of the cone $S^3$ shrinks to zero size but 
$S^2$ shrinks to a finite size.

The second deformation of the orbifold is described by
$$z_1^2+z_2^2+z_3^2=-Cz_4^2 \eqno(8)$$
where $0 \leq C \leq 1$ is a constant. Clearly when $C=1$ we have the conifold. On the other hand, when $C=0$, eq. (8) describes the $Z_2$ ALE space. This means that the 
conifold becomes $ALE \times T^2$ when $C=0$. Thus, eq. (8) describes a one parameter deformation of the direct product $ALE \times T^2$ towards an $ALE$ fibered over an $S^2$.
We parametrize the constant $C$ by $C=sin \theta$; in section 4 we will see the geometrical meaning of this parametrization.

Since we are interested in the conifold deformed by both deformations we consider the six-dimensional space given by
$$z_1^2+z_2^2+z_3^2=-Cz_4^2+R^2 \delta(z_4) \eqno(9)$$
with $C$ and $R$ constants. Note that the last term is nonzero only at the tip of the conifold where we blow up an $S^2$. If this term were nonzero for any $z_4$, it would give us
a deformation of the conifold which describes a blow up of an $S^3$; however, we are not interested in this.

\bigskip
\centerline{\bf 3. D--Brane Inflation on a Resolved and Deformed Conifold}
\medskip

We now consider a D3 brane (along $X_0,X_1,X_2,X_3$ directions) transverse to a six--dimensional compact Calabi--Yau manifold which locally (around the brane) looks like the deformed 
conifold given in eq. (9).
Since the space transverse to the brane is compact, there is four dimensional gravity on the brane world--volume with $M_P^2 \sim V_{CY}/g_s^2 \ell_s^8$ where $V_{CY}$ is the 
volume of the Calabi--Yau manifold. The conifold breaks supersymmetry to $1/4$ and the brane breaks an additional $1/2$; as a result we get ${\cal N}=1$ supersymmetry on the world--volume.
For one such D3 brane, the matter content is given the $U(1) \times U(1)$ gauge group and two pairs of bifundamental chiral multiplets, $A_{1,2}$ an $B_{1,2}$ and a pair of neutral
scalars $\Phi, \bar \Phi$. $A_i$ and $B_i$ carry the charges $(1,-1)$ and $(-1,1)$ respectively[\KLE]. The D3 brane can be broken into two fractional D3 branes under certain conditions.
These fractional D3 branes carry half the charge and tension of a regular D3 brane and can be considered as D5 branes wrapped over a vanishing (or as in our case nonvanishing) $S^2$ at 
the tip of the conifold[\MAL,\BER,\ZAF].
When the D3 brane is at $X_8=X_9=0$ (or at the origin of the corresponding $S^2$), it can be broken into two pieces that may move independently along $X_4,X_5$ (or the other corresponding
$S^2$). The scalars $A_i$ and $B_i$ describe the position of the fractional D3 branes along the $X_6,X_7$ and $X_8,X_9$ directions respectively. $\Phi$ and $\bar \Phi$ parametrize the 
distance between the fractional branes along the $X_4,X_5$ directions.
This theory has the superpotential[\KLE]
$$W(A_i,B_i,\Phi, \bar \Phi)=g~Tr \Phi(A_1B_1+A_2B_2)+ g~Tr \bar \Phi(B_1A_1+B_2A_2)+ m \Phi \bar \Phi \eqno(10)$$
Note that the Yukawa coupling is given by the gauge coupling due to the original ${\cal N}=2$ supersymmetry[\BER] with
$${1 \over g^2}= {1 \over {(2 \pi)^3 g_s \ell_s^2}} \int_{S^2} \sqrt {det(G+B)}={R^2 \over {2 \pi^2 g_s \ell_s^2}} \eqno(11)$$

The scalar mass $m$ in eq. (10) is related to the conifold deformation parameter $C=sin \theta$. When $C=0$, the conifold becomes $ALE \times T^2$ and there is ${\cal N}=2$
supersymmetry on the D3 brane. This means that $m=0$ for $C=0$. On the other hand, when $C=1$, we have the (not deformed) conifold with ${\cal N}=1$ supersymmetry without the scalars
$\Phi, \bar \Phi$ (coming from the vector multiplet). Thus, for $C=1$ we get $m \to \infty$ so that the neutral scalars decouple. These conditions can be satisfied by
$$m_{\Phi}={C \over {(1-C^2)^{1/2}}}{1 \over {2 \pi \ell_s}}= {tan \theta  \over {2 \pi \ell_s}} \eqno(12)$$

The sum of the $U(1)$'s, $(1/2)[U(1)_1+U(1)_2]$ is irrelevant for our purposes ($A_i$ and $B_i$ are neutral under it) and can be neglected. Under the orthogonal combination 
$(1/2)[U(1)_1-U(1)_2]$ the $A_i$ and $B_i$ have charges of $1$ and $-1$ respectively. We can set the $A_2$ and $B_2$ to zero for our purposes. (However, it was argued in ref. [\ACH] that
the presence of these scalars is crucial for eliminating dangerous cosmic strings that form after inflation.)
In addition, we are left with only one
neutral scalar related to the difference of the $U(1)$'s, $(1/2)(\Phi+ \bar \Phi)$. Renaming this neutral scalar $\Phi$ (with an abuse of notation) and $\phi_1=A_1$ and $\phi_2=B_1$
we get the scalar potential (from the F--terms) 
$$V_F(\phi_1,\phi_2,\Phi)=g^2(|\phi_1|^2+ |\phi_2|^2)|\Phi|^2+ g^2 |\phi_1|^2 |\phi_2|^2 +m^2|\Phi|^2 \eqno(13)$$

There is an additional D--term contribution to the scalar potential from the two charged scalars $\phi_{1,2}$. Moreover the resolution of the conifold, i.e. orbifold singularity of the 
ALE fiber at $X_6=X_7=X_8=X_9=0$ (and at $X_4=X_5=0$) by blowing it up by a sphere of radius $R$, correponds to an anomalous D--term, $\xi$, on the brane world--volume[\DGM]. 
The D--term contribution to the scalar potential is
$$V_D=g^2(|\phi_1|^2-|\phi_2|^2+\xi)^2 \eqno(14)$$

The extra tension of the wrapped D5 brane due to the blow--up must be equal to the energy density on the brane world--volume that arises from the resolution of the singularity, 
i.e. the anomalous D--term. Thus we have 
$$T_{D3}=4 \pi R^2 T_{D5}={{4 \pi R^2} \over {(2 \pi)^6 g_s \ell_s^6}}= g^2 \xi^2  \eqno(15) $$ 
which gives
$$\xi={R^2 \over {4 \sqrt 2 \pi^{7/2} g_s \ell_s^4}} \eqno(16)$$

The total scalar potential on the D3 brane world--volume $V_{tot}=V_F+V_D$ is precisely the potential that gives hybrid inflation with the trigger field $\phi_2$ and the inflaton $\Phi$
defined by
$$\Phi={{X_4+iX_5} \over {2 \pi \ell_s^2}} \qquad \phi_1={{X_6+iX_7} \over {2 \pi \ell_s^2}} \qquad \phi_2={{X_8+iX_9} \over {2 \pi \ell_s^2}} \eqno(17)$$
We remind that, for the conifold, the complex planes $X_4+iX_5$ and $X_8+iX_9$ are compactified to two $S^2$'s and the plane $X_6+iX_7$ describes the $U(1)$ fiber and the radial direction. 
Consider an initial state with $\phi_1,\phi_2>>\Phi>g \sqrt \xi$.
In this case, due to their large mass (i.e. $m>H$) $\phi_{1,2}$ will settle to the minimum of their potential at $\phi_1=\phi_2=0$. For $m_{\Phi}<H$, the inflaton, $\Phi$ will slowly
roll down its potential which corresponds to the inflationary era. From the bulk point of view this describes fractional D3 branes slowly moving along the $X_4,X_5$ directions or on the $S^3$
(with both branes at the same $X_6,X_7,X_8,X_9$ coordinates). When the branes get too close to each other and $\Phi^2<g^2 \xi$, the trigger field $\phi_2$ becomes tachyonic and starts to roll
towards its new minimim at $\phi_2=\sqrt \xi$. In other words, as the branes approach each other along $X_4,X_5$ ($S^3$), they start separating along $X_8,X_9$ ($S^2$). Inflation ends when 
the slow roll conditions are violated, i.e. $m_{\phi_2} \sim H$ or $m^2_{\Phi} \sim g^2(-\phi_2^2+\xi)^2/M_P^2$. The final state is given by $\phi_1=0$, $\phi_2=\sqrt \xi$ and $\Phi=0$
which describes two fractional branes at the same $X_4,X_,X_5,X_6,X_7$ coordinates but separated along $X_8,X_9$. Thus at the end, the branes are at the same $r$ and $S^3$ coordinates
and separated only on the blown--up $S^2$.

In order to be realistic, our model of D--brane inflation has to satisfy the WMAP constraints[\MAP]. First and foremost, the slow--roll constraints given by $0<\epsilon_1<0.022$ and 
$-0.06<\epsilon_2<0.05$ have to be satisfied. Here $\epsilon_1=\epsilon$
and $\epsilon_2=2(\epsilon-\eta)$ which are defined by the slow--roll parameters
$$\epsilon={M_P^2 \over 2} \left(V^{\prime} \over V \right)^2 \eqno(18)$$
and
$$\eta=M_P^2 \left(V^{\prime \prime} \over V \right) \eqno(19)$$
Inflation occurs when $\epsilon,\eta <<1$ and ends when at least one of them becomes of $O(1)$.
Our model also has to produce the correct amount of scalar density perturbations; $18.8\times 10^{-10}<A_s<24.8 \times 10^{-10}$ where
$$A_s={H^2 \over {8 \pi^2 M_P^2 \epsilon_1}} \eqno(20)$$
$A_s$ is related to the magnitude of density perturbations by $\delta \rho/\rho \sim A_s^2$.
The ratio of the amplitudes for the tensor and scalar perturbations must satisfy $0<R<0.35$ where $R \sim 16 \epsilon_1$. The scalar and tensor spectral indices which parametrize
the deviations from scale invariant perturbations are constrained by
$0.94<n_s<1.02$ and $-0.044<n_t<0$ where
$$n_s \sim 1-2 \epsilon_1 -\epsilon_2 \qquad \qquad n_t \sim -2\epsilon_2 \eqno(21)$$
In addition, the models must result in about 60 e--folds of inflation
$$N = M_P^{-2} \int {V \over V^{\prime}} d\phi \sim 60 \eqno(22)$$

Using the total potential, $V_{tot}=V_F+V_D$ given by eqs. (13) and (14) we find for the slow--roll conditions (with the parametrization $\Phi=s/2 \pi \ell_s^2$)
$$\epsilon_1={{2M_P^2 m^4 \Phi_i^2} \over {g^4 \xi^4}}={{8 \pi^4 M_P^2 g_s^2 \ell_s^4 tan^4 \theta} \over {R^4}} s^2 \eqno(23)$$  
and
$$\epsilon_2={{4 M_P^2 m^2} \over {g^2 \xi^2}} \left({{m^2 \Phi_i^2} \over {g^2 \xi^2}}-1 \right) = {{64 \pi^5 M_P^2 tan^4 \theta g_s \ell_s^6} \over R^4} \left({{s^2 tan^2 \theta \pi g_s
\ell_s^2} \over R^2}-1 \right)  \eqno(24)$$
The number of e--folds is given by
$$N={{g^2 \xi^2} \over {2 M_P^2 m^2}} log({\Phi_i \over \Phi_c})={R^4 \over {64 \pi^5 M_P^2 g_s tan^2 \theta \ell_s^6}} log({s_i \over s_c}) \eqno(25)$$ 
where $s_i,s_c$ are the initial and critical values of the distance between the branes respectively.
The magnitude of scalar density perturbations is
$$A_s={{g^4 \xi^6} \over {16 \pi^2 M_P^6 m^4 \Phi_i^2}}={{R^8} \over {(32)^3 \pi^{15} g_s^4 \ell_s^{12} M_P^6 s^2 tan^4 \theta}} \eqno(26)$$
The constraints given by eqs. (23)-(26) must be satisfied by the string parameters $\theta, g_s, \ell_s, R$.

\bigskip
\centerline{\bf 4. An Alternative Description}
\medskip

Our D--brane inflation model is closely related to others as we show below. First consider the conifold with $C=0$ in eq. (9). In this case the space transverse to the 
fractional D3 branes becomes $ALE \times T^2$. The neutral scalar mass vanishes and supersymmetry is enhanced to ${\cal N}=2$. This is precisely the model considered in
ref. [\BLO] and results in D--term inflation[\DTE,\BIN,\TYP,] on the fractional branes.

Our model can also be realized in a Hanany--Witten type of brane construction. Unfortunately these cannot be compactified and therefore serve only as a realization of our model
close to the orbifold (or conifold) singularity. In the brane construction the conifold is described by an NS5 brane along the $X_0,X_1,X_2,X_3,X_4,X_5$ directions with another
NS5 brane perpendicular to the first one along the $X_0,X_1,X_2,X_3,X_8,X_9$ directions (and denoted by $NS5_{\pi/2}$ due to the rotation from the $X_4,X_5$ plane to the $X_8,X_9$ plane). 
The branes are at the same $X_7$ coordinate but also separated along 
the $X_6$ direction which is compactified. Compactification of the $X_6$ direction can be described by
two NS5 branes of the first type (which are identified) with an $NS5_{\pi/2}$ brane between them. The fractional D3 branes of our model are now described
by two segments of D4 branes along $X_0,X_1,X_2,X_3,X_6$ and stuck between the two types of NS5 branes. As long as the two NS5 branes are at the same $X_7$ coordinate, the segments
of D4 branes are free to move along $X_4,X_5$. The neutral scalar $\Phi$ describes their position on this plane. The two charged scalars $\phi_1$ and $\phi_2$ parametrize the position
of the segments on the $X_4,X_5$ and $X_8,X_9$ planes respectively. 

The resolution of the conifold by blowing up an $S^2$ at its tip is described by separating the NS5 and $NS5_{\pi/2}$ branes along the $X_7$ direction. As expected, this results in an anomolous
D--term on the D4 brane world--volume. The deformation given by $C$ in eq. (8) is described by the rotation of one of the NS5 branes relative to the other (so that the angle between them
becomes $\pi/2-\theta$). This makes sense because if we make the two NS5 branes parallel (i.e. $\theta=\pi/2$) then we get an ${\cal N}=2$ theory which corresponds to an $ALE \times T^2$
as required. Thus, in this case, the deformation is described by a physical rotation in the bulk by an angle $\theta$. The mass of $\Phi$ is given by $m_{\Phi}=tan \theta/ 2\pi \ell_s$
which is exactly eq. (12). This justifies our parametrization of the parameter $C$ in terms of an angle. 

The brane construction described in the previous paragraph also gives the scalar potential in eqs. (13) and (14) and results in hybrid inflation on the D4 branes (with the caveat about 
compactification). The two fractional branes (segments) are initially separated along $X_4,X_5$. Due to the tilt of the $NS_{\theta}$, this is not a supersymmetric configuration. As a
result, the two segments move slowly toward each other which gives hybrid inflation. In the meantime, the segments separate along the $X_8,X_9$ directions. The final state is given by the
segments separated only along the $X_8,X_9$ directions.

This is quite similar to the brane constructions in refs. [\KAL] and [\LAS]. However, note that the D6 brane that appears in those cases is missing above. In this model the scalar fields
arise from strings stretched between the two fractional D4 branes (segments) rather than between a D4 and a D6 brane. The absence of the D6 is crucial for the compactification
of the transverse space.

\bigskip
\centerline{\bf 5. Conclusions and Discussion}
\medskip

In this letter, we obtained hybrid inflation on fractional branes transverse to a resolved and deformed conifold. The origin of inflation is the resolution of the conifold by the
blow--up of its tip. The inflaton which describes the slow motion of the two fractional D3 branes on the $S^3$ obtains its mass from the deformation of the conifold given by
eq. (8). By deriving a brane/bulk dictionary of the parameters we showed that the model can satisfy the WMAP constraints. We also discussed the relation of our model to other
D--brane inflation models, most notably those that are built as Hanany--Witten models.

It is well--known that, after inflation, when the scalars reach their minimum (with potential in eqs. (13) and (14)) cosmic strings will be created[\TYE,\POL]. These are not the recently
discovered cosmic D--term strings[\COS,\STR] even though we can have D3 branes wrapping the resolved $S^2$ at the tip of the conifold. The reason is the existence of the scalar potential which
must vanish for the D--term strings to be stable[\COS]. In the string setup this can only occur if the transverse manifold is a conifold which is not deformed, i.e. when $C=0$. The cosmic
strings that are generated would seem to be dangerous since they will contribute a large amount to the energy density and result in density fluctuations that violate the WMAP bounds.
However, recently it was noted that the existence of the second pair of charged scalars (that arise from the underlying ${\cal N}=2$ supersymmetry of the model and have been set to zero)
solves this problem[\ACH]. Due to this extra pair, the cosmic strings are not local but only semilocal. They are marginally stable since the Yukawa coupling equals the gauge coupling[\ACH]. 
As a result, these cosmic strings are not created after inflation. 

The model can also describe quintessence on a brane with another (but unatural) choice of parameters. Taking $\xi \sim 10^{-60} M_P^2$ would give a reasonable model of quintessence
without explaining the smallness of the vacuum energy. The above value of $\xi$ requires an unnaturally small value for the blow--up radius. Note that this case would correspond to
hybrid quintessence[\QUI] which ends after a finite time and therefore does not suffer from problems related to string theory on space--times with cosmological horizons.

It would also be interesting to find a holographic description of D--brane inflation of the fractional branes along the lines of the dS/CFT correspondence[\STR]. Inflation on the brane
would correspond to a renormalization group flow of a ${\cal N}=1$, Euclidean 3D theory towards a fixed point[\INF,\HOL]. A dictionary between the Euclidean CFT and the bulk physics 
would help to clarify these issues.

Our model can be easily generalized by considering a cone with a base that is given by an $Z_n$ ALE space fibered over $S^2$. In this case, the brane theory becomes a $U(1)^n$ quiver
theory with mater in the bifundamental representations. There are $n-1$ possible $S^2$ that can be
blown up at the tip of the cone. Each one of these would correspond to a different D--term on the brane world--volume and would be a source of inflation. As above, the inflaton mass
arises from the deformation of the fibration. The D--brane inflation scenario in the bulk and hybrid inflation on the brane would occur exactly as in our model. Thus, it seems that
our model is quite robust and describes inflation on many singularities of Calabi--Yau manifolds in addition to the conifold.



\vfill

\refout

\end
\bye